\documentclass[twocolumn,showpacs,prb,preprintnumbers,amsmath,amssymb,floatfix]{revtex4}

\usepackage{epsfig,psfrag}
\usepackage{dcolumn}
\usepackage{bm}
\usepackage{graphicx}
\usepackage{color}

\begin{document}

\title{Superconducting nonequilibrium transport through a weakly interacting
quantum dot}
\author{L.~Dell'Anna, A.~Zazunov, and R.~Egger}
\affiliation{Institut f\"ur Theoretische Physik, Heinrich-Heine-Universit\"at,
D-40225  D\"usseldorf, Germany}
\date{\today}

\begin{abstract}
We study the out-of-equilibrium current through an interacting
quantum dot modelled as an Anderson impurity contacted by
two BCS superconductors held at fixed voltage bias.
In order to account for multiple Andreev reflections, we develop a
Keldysh Green's function scheme perturbative
in the dot's interaction strength.  We find an unexpected enhancement of the
current due to repulsive interactions for small to intermediate lead-to-dot couplings.
\end{abstract}

\maketitle

\section{Introduction}

Superconducting transport through low-dimensional nanoscale structures
is currently attracting considerable interest.  Gate-tunable Josephson
currents through nanowire-based quantum dots have been reported,\cite{vandam}
and similar setups have been realized using (short)
carbon nanotubes\cite{tube,buitelaar} and
metallofullerene molecules.\cite{fulleren}
Nonequilibrium transport in such systems contacted
by superconducting electrodes has been a particular focus of recent
experimental effort,\cite{buitelaar,fulleren,jorgensen,lieber,jarillo,christian,lindelof}
mainly caused by an interesting interplay between interaction
effects (on the quantum dot) and superconducting correlations
(due to the electrodes).
One remarkable consequence is the observation of an
`even-odd' effect (as a function of the dot's occupation number)
in the conductance.\cite{christian,lindelof}
This effect is presumably caused by the absence or presence of Kondo correlations.
In this paper, we analyze superconducting transport through an interacting
quantum dot for the simplest case of a
single spin-degenerate level with repulsive on-site interaction energy $U>0$
(Anderson impurity), contacted by two wide $s$-wave BCS superconducting electrodes
with identical gap $\Delta$.
For simplicity, we assume that both lead-to-dot couplings (hybridizations) are equal,
$\Gamma_L=\Gamma_R=\Gamma$, and consider
the two electrodes held at potential difference (voltage bias) $V$.
Under a systematic perturbative expansion in the interaction strength $U$,
we compute the $I-V$ characteristics, in particular for the interesting
subgap regime $eV<2\Delta$, where multiple Andreev reflection (MAR)
processes provide the dominant transport mechanism.
A theory of coherent MAR has been originally developed for
superconducting point contacts,\cite{mar}
with the essential assumption that charging interaction effects
inside the contact can be neglected.
The problem of resonant MAR through a noninteracting quantum level has been treated
in Refs.~\onlinecite{ingerman,cuevas,wendin}.

While the interacting problem in equilibrium has been
theoretically studied by many authors,\cite{siano}
the corresponding nonequilibrium problem is more difficult and
far less understood.  Previous approaches can be broadly grouped in three classes.
(i) By ignoring MAR processes in the Coulomb blockade regime,
additional side-peaks in the differential conductance
at $eV=2 \Delta$ and $2(\Delta + U)$ were predicted,\cite{kang}
reflecting the singularity of the BCS spectral density of the leads.
(ii) Different mean-field schemes have been proposed,\cite{cuevas,avishai,slave}
based on slave-boson or Hubbard-Stratonovich-path-integral approaches.
These calculations predict an overall suppression of the current by the
interactions. This suppression is obtained only
for sufficiently repulsive interactions, while there
is no interaction effect for weak interaction.\cite{avishai}
(iii) A Fermi liquid approach
valid in the deep Kondo limit has been proposed.\cite{alfredo}
Here we do not discuss the Kondo regime, but instead focus on the
limit of weak interactions, $U/ \Gamma<1$, where a
controlled perturbative expansion in the small parameter $U/\Gamma$
is possible. Note that this approach still allows for arbitrary
ratio $\Gamma/\Delta$. Such calculations have been carried out
for normal-conducting ($\Delta=0$) electrodes recently,\cite{normal,hamasaki,eg}
and we here generalize them to superconducting electrodes.
The case $U< \Gamma$ is of experimental relevance for
the understanding of superconducting transport through quantum dots
or molecules with good lead-to-dot couplings.

The structure of the remainder of this article is as follows.
In Sec.~\ref{sec2}, we discuss our perturbation theory approach
to superconducting transport through an Anderson dot, and its
numerical implementation.  Results for the current-voltage
characteristics are shown and discussed in Sec.~\ref{sec3}.
The appendix contains qualitative arguments for the
current enhancement found at $\Gamma<\Delta$,  based on an
evaluation of the Josephson current. We often set $\hbar=e=1$.

\section{Perturbative approach to superconducting transport}
\label{sec2}

We consider the canonical Anderson impurity model, $H=H_D+H_T
+H_L+H_R$, where a single-level dot with spinful fermion $d_\sigma$ ($H_D$) is
tunnel-coupled ($H_T$) to left/right superconducting reservoirs $H_{L/R}$
held at chemical potential difference $eV$. The isolated dot corresponds to
($n_\sigma=d^\dagger_\sigma d^{}_\sigma=0,1$)
\begin{equation} \label{model}
H_D=  E_0(n_\uparrow +n_\downarrow) + U n_\uparrow n_\downarrow
= \epsilon_0 (n_\uparrow+n_\downarrow)
-\frac{U}{2}(n_\uparrow - n_\downarrow)^2   .
\end{equation}
The `noninteracting' model below is taken to contain the interaction level
shift $\epsilon_0=E_0+U/2$ of the bare level $E_0$.
The leads are described by a pair of $s$-wave BCS Hamiltonians
in the standard wide-band limit.
We are interested in the $V\ne 0$ case,
and take the same real-valued gap parameter $\Delta>0$ for both electrodes.
Using the Nambu vector $\Psi^T_{j,k}=(\psi_{j,k,\uparrow},
\psi^\dagger_{j,-k,\downarrow})$ for electrons in lead $j=L/R$, we thus have
(we put $\hbar=e=1$ in intermediate steps)
\begin{equation} \label{h2}
H_{j} = \sum_k \Psi_{jk}^\dagger \left (
 (k^2/2m -\epsilon_F)\sigma_z +  \Delta  \sigma_x \right) \Psi_{jk},
\end{equation}
with Pauli matrices $\sigma_{i}$  ($\tau_{i}$)
 in Nambu (Keldysh) space.
Using the Nambu vector $d=(d^{}_\uparrow, d^\dagger_\downarrow)^T$
and $\Gamma=\pi\nu_0 |t_0|^2$ for
(normal) lead density of states $\nu_0$, the lead-dot coupling is
\begin{equation}\label{h3}
H_T = t_0 \sum_{k,j=L/R=\pm} \Psi^\dagger_{jk}
\sigma_z e^{\pm i\sigma_z V t/2} d + {\rm h.c.},
\end{equation}
where the voltage $V$ enters via the time-dependent phase.
We now define the Keldysh-Nambu Green's function for the dot fermion,
\begin{equation}\label{green}
G^{ss'}_{\alpha\alpha'}(t,t') = -i\langle {\hat T}_C [d_\alpha (t_s)
d^\dagger_{\alpha'}(t_{s'}) ]\rangle,
\end{equation}
where $\alpha,\alpha'=1,2$ ($s,s'=1,2$) are Nambu (Keldysh) indices,
and $\hat T_C$ is the time-ordering operator along the Keldysh
contour. Accordingly, $t_s$ denotes a time
taken on branch $s$ of the Keldysh contour.
It is convenient to use the Fourier decomposition \cite{zazu}
\begin{equation}\label{greennew}
G^{ab}(t,t') = \sum_{n, m = -\infty}^{\infty}
\int_F \frac{d \omega}{2 \pi} \,
e^{-i \omega_n t + i \omega_m t'} \, G^{ab}_{nm}(\omega) ,
\end{equation}
where $a,b=1,\ldots,4$ denotes Nambu-Keldysh indices defined
by $a=\alpha+2(s-1)$, and $\omega_n = \omega + n V$
for $\omega$ within the 'fundamental' domain
$F\equiv [-V/2,V/2]$.
For fixed $\omega\in F$, the Dyson equation
for the full Green's function $\check{G}$
(the check notation refers to the Keldysh-Nambu structure),
 \begin{equation}\label{g1}
\check{G}^{-1} = \check{G}_{0}^{-1} - \check{\Sigma},
\end{equation}
then becomes a matrix equation suitable for numerical inversion. Here,
interaction effects are encoded in the self-energy $\check{\Sigma}$.
After integrating out the lead fermion degrees of freedom,
the noninteracting Green's function $\check{G}_0$ is
\begin{equation} \label{g0}
\check{G}_{0,nm}^{-1}(\omega) =
(\omega_n-\epsilon_0 \sigma_z) \tau_z  \delta_{nm}
- \Gamma \sum_{j=L/R} {\check \gamma}_{j,nm}(\omega).
\end{equation}
The self-energy due to tracing out the respective lead
is given by the Nambu matrix
\begin{eqnarray} \label{selfen}
&& \check{\gamma}_{j=L/R=\pm,nm}(\omega) = \\ \nonumber
&& \left( \begin{array}{cc}
\delta_{nm} \, \check{X}(\omega_n \mp V/2) &
\delta_{m, n \mp 1} \, \check{Y}(\omega_n \mp V/2) \\
\delta_{m, n \pm 1} \, \check{Y}(\omega_n \pm  V/2) &
\delta_{nm} \, \check{X}(\omega_n \pm V/2)
\end{array} \right)
\end{eqnarray}
with Keldysh matrices $\check{Y}(\omega) = - \Delta\check{X}(\omega)/\omega$
and
\[
\check{X}(\omega) =   \left\{ \begin{array}{ll}
- \frac{\omega}{\sqrt{\Delta^2 - \omega^2}} \tau_z ,
 & |\omega|<\Delta\\
 \frac{i|\omega|}{\sqrt{\omega^2-\Delta^2}} \left( \begin{array}{cc}
2 f_\omega - 1 & - 2 f_\omega \\ 2 f_{-\omega} & 2 f_\omega - 1
\end{array} \right) , & |\omega|>\Delta \end{array}\right.
\]
where $f_\omega=1/(1+e^{\omega/k_B T})$ is the Fermi function.
The steady-state dc current through the left/right junction
then follows as
\begin{equation} \label{dc_current}
I_{L/R} = \mp  2\Gamma\ {\rm Re} \sum_{nm} \int_F \frac{d \omega}{2 \pi}
{\rm tr}\left[ \sigma_z\check{\gamma}_{L/R,nm}(\omega)
 \check{G}_{mn}(\omega) \right]^{+-},
\end{equation}
where the trace is only over Nambu space, and
$(+-)$ refers to the $(12)$ Keldysh component.
Current conservation, $I_L=I_R=I$, is fulfilled for all results below.

\subsection{First order}

\begin{figure}[h!]
\scalebox{0.25}{\includegraphics{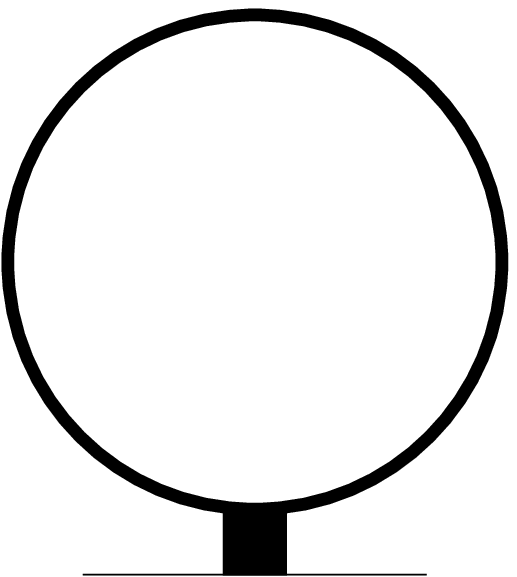}}
\hspace{1cm}
\scalebox{0.25}{\includegraphics{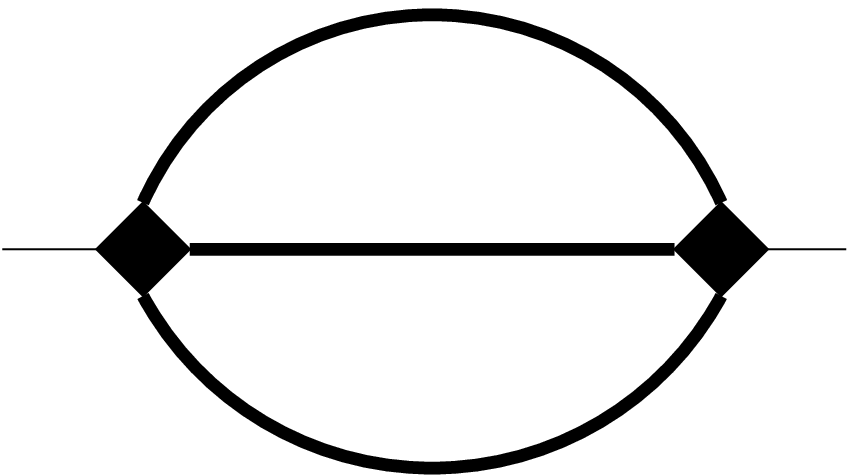}}
\caption{ \label{f1} Electron-electron interaction
self-energy diagrams taken into account in this paper:
(a) First order (left). (b) Second order (right).  }
\end{figure}

Since the exact self-energy $\check{\Sigma}$ is not known, we proceed in a
perturbative fashion, starting with the first-order self-energy
in Fig.~\ref{f1}(a), made self-consistent by using the
full $\check{G}$ in the diagram.
It is convenient to introduce the four-point vertex, cp.~also
Ref.~\onlinecite{normal},
\begin{equation}\label{lambda}
\Lambda_{abcd} = \frac{1}{2}\Big(\delta_{ab}\tau_{cd}+\delta_{cd}\tau_{ab}-\delta_{ad}\tau_{cb}
-\delta_{cb}\tau_{ad}\Big),
\end{equation}
where $\tau=\sigma_0 \tau_z = {\rm diag}(1,1,-1,-1)$. Only 8 out
 of the possible 256 entries of the tensor $\Lambda$ are nonzero,
 with values $\pm 1$.
With this convention, the  complete first-order self-energy is given by
\begin{equation}\label{self1}
\Sigma^{(1) ab}_{n,n+m} =
i\, {U} \Lambda_{abcd} \sum_{n'} \int_F\frac{d\omega'}{2\pi}
\, G^{dc}_{n',n'+m}(\omega'),
\end{equation}
where we use the sum convention for $c,d$.
Note that this self-energy is independent of $(n,\omega)$, but still
depends on the `off-diagonal' frequency index $m$.
In time representation, $m\ne 0$ contributions come with phase factors
$e^{\pm 2im V t}$ and thus correspond to anomalous (pairing) correlations.
 The presence of the off-diagonal harmonics in the self-energy is
a consequence of the coherent MAR transport regime considered here.
The role of coherence is particularly important
for the interplay between MAR processes and charging effects in a quantum dot
with a relatively strong coupling to the leads.

Let us at that stage briefly compare the first-order
self-consistent approach based on Eqs.~(\ref{g1}) and (\ref{self1})
to the  mean-field approximation of Ref.~\onlinecite{avishai}.
The latter effectively considers only the time averaged components
of  the self-energy (\ref{self1}) corresponding to the two first terms of Eq.~(\ref{lambda}),
thereby discarding all harmonics with $m\ne 0$ and exchange terms in Eq.~(\ref{self1}).
The resulting self-energy contributions taken into account in
 Ref.~\onlinecite{avishai} correspond to $\gamma_+\tau_0$ and
$\gamma_-\tau_z$. The scalar constants $\gamma_\pm$
 can be written as
\begin{eqnarray}\nonumber
\gamma_+ &=& i\frac{U}{2}\sum_{n'}
\int_F\frac{d\omega'}{2\pi}\,{\rm Tr} \left(
\tau_z \check{G}_{n' n'}(\omega') \right),
\\ \label{zaikin1}
\gamma_-&=& i\frac{U}{2}\sum_{n'} \int_F\frac{d\omega'}{2\pi}
\,{\rm Tr}\left( \check{G}_{n'n'}(\omega')\right),
\end{eqnarray}
where the trace is over both Nambu and Keldysh spaces.
Under this approximation, there is no interaction effect on the
current $I(V)$ below some critical value $U_c$. In fact,
 nontrivial stable solutions $\gamma_\pm\ne 0$
 for the self-consistency equation
(\ref{zaikin1}) exist only for $U>U_c$,\cite{avishai} where $U_c$
depends on $V,\Gamma$ and $\Delta$.
The symmetry-broken phase with $\gamma_-\ne 0$ corresponds to
a spin-polarized dot, and one then finds a Coulomb blockade
suppression of the current.\cite{avishai}
However, for normal leads, serious problems with spin-polarized
out-of-equilibrium mean-field solutions
for the Anderson dot have been identified recently,\cite{zarand}
and those arguments also apply to the superconducting case.
A typical value is $U_c\approx\Gamma$ for $V\approx \Gamma \approx \Delta/2$,
and we shall always limit ourselves  to $U<U_c$
where no such problems arise.
In our calculations, the actual value of $U_c$ follows from the numerical
solution of the self-consistency problem, and we can thereby ensure that
no spin-polarized solutions are present.  In contrast to the mean-field
scheme of Ref.~\onlinecite{avishai}, however, the full
first-order self-consistent
approach generates sizeable interaction corrections even for
small $U$, see Sec.~\ref{sec3}.  These corrections are not
just a matter of numerical accuracy but reflect the importance
of on-dot pairing terms.

\subsection{Second order}

To go beyond the self-consistent first-order approximation
given by Eq.~(\ref{self1}), we have to evaluate the
second-order diagram shown in Fig.~\ref{f1}(b).
Due to the large numerical effort in evaluating this diagram, we restrict
ourselves to a non-selfconsistent scheme at this point, i.e., we use the
solution $\check{G}$ of the first-order problem to evaluate
 $\check{\Sigma}^{(2)}$.
As is well known,\cite{baym,hershfield} under such a scheme
 current conservation is only ensured
for the particle-hole symmetric case, $\epsilon_0=0$, see also
Ref.~\onlinecite{zazu}.  We therefore show second-order
results only for $\epsilon_0=0$.
The second-order Nambu-Keldysh self-energy reads in time representation
(sum convention)
\begin{eqnarray} \label{self2}
\Sigma^{(2) ab}(t,t') &=& \frac{U^2}{2}
\Lambda_{afgh} \Lambda_{ebcd}  \\ \nonumber &\times&
G^{fe}(t,t') G^{dg}(t',t) G^{hc}(t,t').
\end{eqnarray}
To avoid numerically expensive frequency convolutions, it is convenient to
first compute the Green's function in time representation
according to Eq.~(\ref{greennew}),
then evaluate the self-energy in Eq.~(\ref{self2}), and
finally transform this result back to frequency
space to use it in the Dyson equation (\ref{g1}).
Notice that Eq.~(\ref{self2}) corresponding to the skeleton
diagram in Fig.~\ref{f1}(b) represents only a part
of all possible second-order contributions.
The rest is given by the time-local piece
\[
-U^2\Lambda_{abcd}
\Lambda_{efgh}\int dt^{\prime\prime}\,{G}^{de}(t,t^{\prime\prime})\,
{G}^{hg}(t^{\prime\prime},t^{\prime\prime})\,
{G}^{fc}(t^{\prime\prime},t) ,
\]
which has already been taken into account by our self-consistent
first-order solution.

\subsection{Calculation of current}

The numerical implementation of the above perturbative approach is straightforward.
To evaluate the current $I(V)$ from Eq.~(\ref{dc_current}),
we partition the frequency summations into windows of width $V$, and impose
a bandwidth cutoff $\omega_c$, such
that $|\omega_n| < \omega_c$.  In our calculations, we use $\omega_c=10\Delta$,
but the precise choice is not critical.
We then discretize the fundamental frequency domain $F$ with a step-size $\delta\omega$,
and use a
fast Fourier transform routine to switch between time and frequency
representations.  (The efficient evaluation of the second-order self-energy
requires to employ the time representation, while the Dyson equation needs
the frequency representation.)
Typically, we found $\delta \omega = 0.005 \Delta$
 sufficient for convergence. The matrix inversion in Eq.~(\ref{g1}) is then performed
for each $\omega \in F$ separately,
involving matrix dimensions of the order of $\omega_c/|V|$.
We refer to Ref.~\onlinecite{zazu} for further details of the numerical
implementation in the related case of a phonon-mediated interaction.

In a first step, we solve the first-order self-consistent
problem posed by Eqs.~(\ref{g1}), (\ref{g0}) and  (\ref{self1}).
This solution proceeds iteratively, where the stability or instability
of the solution for $\check{G}$ is checked carefully by probing small deviations
around it. The iterative solution can in fact
be carried out with very modest computational effort,
and quickly converges to a unique solution (as long as $U<U_c$).
In a second step, we then use this converged first-order
Green's function to evaluate
$\check{\Sigma}^{(2)}$ according to Eq.~(\ref{self2}),
and to finally compute the current from Eq.~(\ref{dc_current}).
 The second-order calculation is
quite time-consuming for low bias voltage,
where many MAR orders need to be taken into account, and we
therefore restricted our
calculations to $eV/\Delta\geq 0.2$.  As consistency
check for our numerical code, we have reproduced known results
for $U=0$, see Refs.~\onlinecite{ingerman,cuevas,wendin},
and the corresponding perturbative-in-$U$ results for normal-conducting
leads ($\Delta=0$),
see Refs.~\onlinecite{normal,hamasaki}.
We have also reproduced the respective
results of Ref.~\onlinecite{avishai} when
implementing their approximations.
As additional check, Green's function sum rules,
such as ${\rm tr}
 \left[ \tau_z \sigma_z \check{G}(t,t) \right] = 0$ at coinciding times,
have been verified.

\section{Results and discussion}\label{sec3}

Next we discuss numerical results obtained
under the perturbative approach described in Sec.~\ref{sec2}.
All results are for $T=0$, and unless noted otherwise, we put $U/\Gamma=0.5$
which is sufficiently small to ensure $U<U_c$ for all investigated $V/\Delta$
and $\Gamma/\Delta$ but large enough to produce significant interaction
corrections to the $I-V$ characteristics.
We will focus on the most interesting subgap regime, $eV<2\Delta$.
The excess current $I_{exc}=\lim_{V\to\infty} [I(V,\Delta)-I(V,\Delta=0)]$
has also been computed.   The interaction contribution $\delta I_{exc}$
to this quantity turns out to be generally small, similar to what
is found for the case of phonon-mediated interactions.\cite{zazu}
Remarkably, this interaction correction is positive for $\Gamma\alt
\Delta$, pointing towards a current enhancement.  This trend is
quite generic and discussed next.

\begin{figure}[ht!]
\scalebox{0.66}{\includegraphics{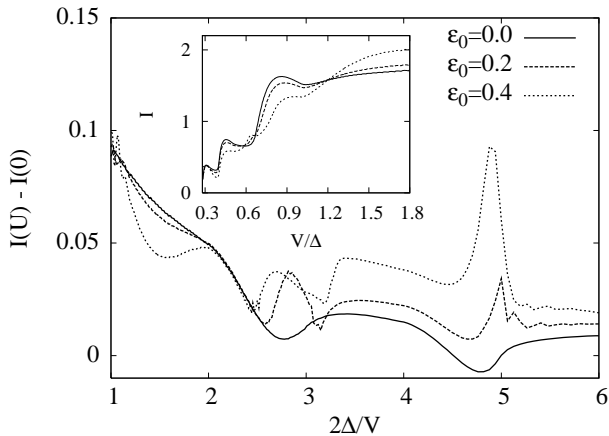}}
\caption{\label{f2}
Interaction correction to the current (currents are always plotted
in units of $e\Delta/h$) from the self-consistent first-order
approach, for $U/\Gamma=\Gamma/\Delta=0.5$ and various $\epsilon_0/\Delta$.
The inverse voltage scale is taken to compare with standard MAR features.
Inset: Full $I-V$ curves for same parameters.}
\end{figure}

\begin{figure}[ht!]
\scalebox{0.65}{\includegraphics{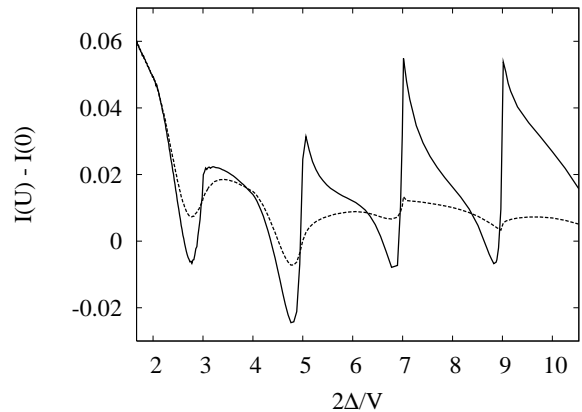}}
\caption{\label{f3}
Same as Fig.~\ref{f2} but for $\epsilon_0=0$. The dashed
curve gives the first-order self-consistent result, while
the solid curve includes also the second-order contribution.}
\end{figure}

\begin{figure}[ht!]
\scalebox{0.65}{\includegraphics{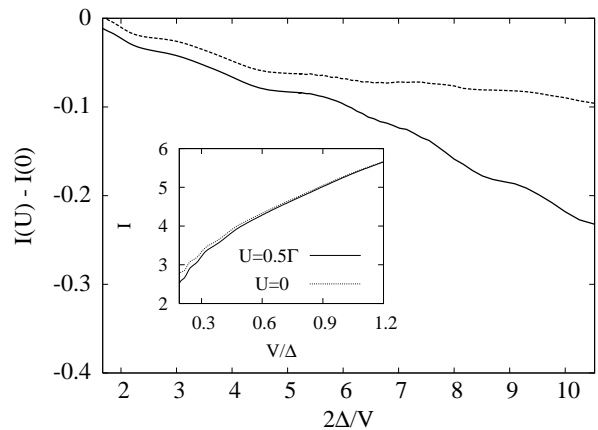}}
\caption{\label{f4}
Same as Fig.~\ref{f3} but for $\Gamma=2\Delta$.
Inset: $I-V$ curve from second-order perturbation theory and for $U=0$.}
\end{figure}

Let us start by showing results obtained from the first-order self-consistent
scheme (i.e., without the second-order self-energy). In that case, by
virtue of self-consistency, we have
the freedom to vary $\epsilon_0$ without spoiling current
 conservation.\cite{baym,hershfield}
Representative numerical results
for the voltage-dependent interaction correction to the current, $I(U)-I(U=0)$,
are shown for $\Gamma/\Delta=0.5$ in Fig.~\ref{f2}.
For all results shown here, we have  $U<U_c$, and
 the approach of Ref.~\onlinecite{avishai}
would not yield any interaction correction.
However, we find significant interaction effects for $U<U_c$
within the full first-order self-consistent approach.
These effects are due to the time-dependent ($m\ne 0$)
parts of the self-energy (\ref{self1}), which contain
 pairing order parameters on the dot.
For instance, at the symmetric point $\epsilon_0=0$, we
 find by perturbation theory in $U$ that $\gamma_\pm=0$, but the
complex-valued $m=1$ pairing term
\begin{equation}\label{delta}
\delta \equiv \Sigma_{n,n+1}^{(1),12}
\end{equation}
stays finite.  This $\omega$-independent off-diagonal Nambu component of
the self-energy, absent in the normal ($\Delta=0$)
case,  describes the effect of
interactions on the proximity-induced pairing
correlation on the dot. At the mean-field level, $\gamma_-$
is dominant for large $U$, while terms like $\delta$ dominate for small $U$.
Similar contributions with $|m|>1$ exist and are kept
in our self-consistent first-order calculations, but they turn out to be
significantly smaller.

Quite remarkably, we find $I(U)>I(U=0)$
for most voltages and/or dot level energies $\epsilon_0$, pointing
to an {\sl enhancement of the MAR-mediated current by repulsive interactions}.
We have persistently found this
unexpected feature throughout the parameter regime
$\Gamma<\Delta$, also when including the second-order self-energy,
see below.  A similar (but weaker) enhancement can be
found analytically for the critical Josephson current
of this system in equilibrium, see Appendix.
The current enhancement is reminiscent yet different from
the `antiblockade' behavior due to dynamical Coulomb blockade
effects on MAR transport discussed in Ref.~\onlinecite{alf3}.
It is also consistent with the crossover
from current enhancement to decrease with growing $\Gamma$
for phonon-mediated interactions and normal-conducting leads.\cite{eg}

\begin{figure}[ht!]
\scalebox{0.65}{\includegraphics{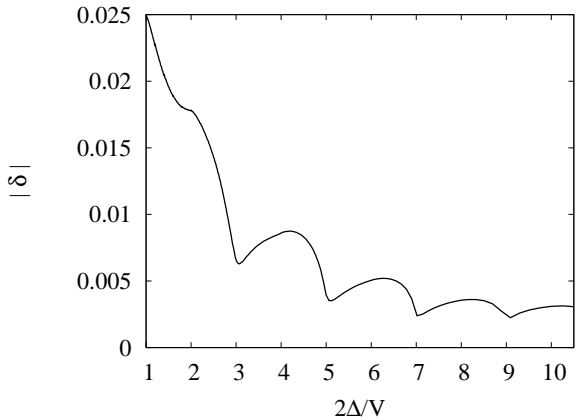}}
\caption{\label{f5} Absolute value of $\delta$, see Eq.~(\ref{delta}),
in units of $\Delta$, for the parameters in Fig.~\ref{f3}.}
\end{figure}

To illustrate the role of the second-order contribution for $U/\Gamma=0.5$,
we now focus on the symmetric case $\epsilon_0=0$, first taking again
 $\Gamma/\Delta=0.5$. The results of the first- and second-order calculations
are compared in Fig.~\ref{f3}.
Notice that the second-order correction, which is the leading
time-nonlocal term in the perturbative expansion, becomes more
and  more important when lowering the voltage.
In agreement with the conclusion drawn from the
first-order self-consistent calculation shown in Fig.~\ref{f2},
a clear enhancement
of the current by interactions can be observed
for a broad range of voltages. This enhancement is especially pronounced
for voltages slightly below the odd MAR
 peaks located at $eV =2\Delta/(2n+1)$.
As indicated by our results for larger $\Gamma$, see Fig.~\ref{f4} for
the case $\Gamma/\Delta=2$, the interaction-induced enhancement of the
current is restricted to small $\Gamma/\Delta$.
For larger $\Gamma/\Delta$, the
current instead is weakly suppressed by interactions.
The same pronounced
MAR peak structure as in Fig.~\ref{f3} can be observed
in the pair order parameter
$\delta$ defined in Eq.~(\ref{delta}), whose absolute value is shown in Fig.~\ref{f5}.
The fact that the characteristic MAR features still appear at $eV=2\Delta/n$ for
the interacting case indicates that, at least for small $U$,
the number of Andreev reflections is not affected by the Coulomb interaction.
This is
in contrast to the inelastic MAR picture for phonon-mediated interactions,
as discussed in Ref.~\onlinecite{zazu}.
Moreover, our results indicate that
the Andreev quasiresonances\cite{ingerman,wendin} are not
shifted away from the gap subharmonics.

As follows from our numerical analysis,
the magnitude of the difference current is mainly determined by
the first harmonics of interaction-mediated pairing, Eq. (\ref{delta}),
and  can be roughly approximated as
\begin{equation}
I(U)-I(0) \sim (e / \hbar) \, | \delta| \Delta / \Gamma ~.
\label{estimation}
\end{equation}
This can be seen, for instance, from comparison of the curves in
Figs.~\ref{f5}  and \ref{f2} (for $\epsilon_0 = 0$).
A similar expression for the difference current, Eq. (\ref{estimation}),
is also obtained from a simple Fermi golden rule 
calculation, by analyzing the dynamics of
Andreev (subgap) states at very low voltages, $e V \ll \Delta$.
The corresponding correction to the transition rate from Andreev
states into the continuum is determined by the imaginary part of the
Andreev state self-energy $\Sigma_{A}(\omega) = \delta
\sqrt{\Delta^2 - \omega^2}/\Gamma$,
while the total escape probability leading to the difference current
is given by an expression similar to Eq. (3) in Ref.~\onlinecite{alfredo}.

\begin{figure}[ht!]
\scalebox{0.65}{\includegraphics{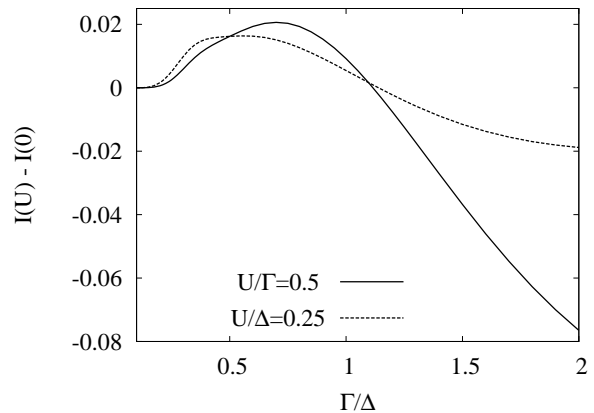}}
\caption{\label{f6}
 $I(U)-I(0)$ versus $\Gamma/\Delta$ at $V=0.6\Delta$,
$\epsilon_0=0$,
fixing $U/\Gamma=0.5$ (solid line) or $U/\Delta=0.25$ (dashed line),
from self-consistent first-order perturbation theory.}
\end{figure}

The above results also suggest that as a function of the ratio $\Gamma/\Delta$,
there should be a crossover from enhancement to suppression of
the current around $\Gamma/\Delta\approx 1$.
This is what we get when fixing the voltage and changing $\Gamma/\Delta$.
In Fig.~\ref{f6}, we have chosen $V=0.6\Delta$, where the self-consistent
first-order approximation gives the main contribution, and plotted
$I(U)-I(0)$ either for fixed $U/\Gamma=0.5$ (solid line), or for
fixed $U/\Delta=0.25$ (dashed line).
The two curves both cross zero approximately at the same value, $\Gamma
\simeq \Delta$.

In conclusion, we have presented a theory exploring
 the effect of weak interactions
on superconducting transport through a quantum dot.
Employing second-order perturbation theory,  valid for $U<\Gamma$,
we find  an unexpected enhancement of the subgap current against
its noninteracting value when the hybridization $\Gamma$
is smaller than the BCS gap parameter $\Delta$.
The perturbation theory scheme pursued in this paper offers
controlled results in one corner of the parameter regime,
and in contrast to previous mean-field theories, we predict
significant interaction corrections even for weak interactions.

\acknowledgments
We thank A. Levy Yeyati, T. Martin, and V. Shumeiko for discussions.
This work was supported by the EU HYSWITCH and INSTANS networks.

\begin{appendix}
\section{Josephson current}

In this appendix, we briefly show that the interaction-induced
 current enhancement found in the $I-V$ curves for $\Gamma<\Delta$
discussed in Sec.~\ref{sec3} also appears in the equilibrium
Josephson current-phase relation (where $\phi$ is the phase difference
across the dot) for the same model.
We consider the corresponding first-order self-consistent theory
in equilibrium, for simplicity at $\epsilon_0=0$ only.
For small $U$, no polarization is present, $\gamma=0$, and
only the proximity-induced mean-field parameter $\delta=U\langle
d^\dagger_\uparrow d^\dagger_\downarrow\rangle$ gives an
effect. Assuming real-valued $\delta$,
the $T=0$ self-consistency equation reads
\begin{equation}\label{selfa}
\delta=U\int_{-\infty}^{\infty} \frac{d\omega}{2\pi}
\frac{\beta_\omega-\delta}{\alpha_\omega^2+
\left(\beta_\omega-\delta\right)^2}.
\end{equation}
where
\[
\alpha_\omega=\omega\left(1+\frac{\Gamma}{\sqrt{\omega^2+\Delta^2}}\right),
\quad \beta_\omega=\frac{\Gamma\Delta\cos(\phi/2)}{\sqrt{\omega^2+\Delta^2}}.
\]
The Josephson current is then given as
\begin{equation}\label{jos}
I =\frac{e\Gamma\Delta}{\pi\hbar}\sin(\phi/2)
\int_{-\infty}^{\infty} \frac{d\omega}{\sqrt{\omega^2+\Delta^2}}
\frac{\beta_\omega-\delta}{\alpha_\omega^2+\left(\beta_\omega-\delta\right)^2}
\end{equation}
The presence of $\delta$ in Eq.~(\ref{jos})
 generally causes two counteracting effects:  there is a
decrease of $I$ due to the numerator, but an increase due to the
appearance of $\delta$ in the denominator.
Which of these is more important can only be clarified by detailed
calculation.
We present analytical evaluations valid for
$|\delta|\ll \Gamma|\cos(\phi/2)|$, separately for the regimes $\Gamma /\Delta
\ll 1$ and $\Gamma/\Delta\gg 1$.

Let us first discuss  $\Gamma/\Delta\ll 1$, where
Eq.~(\ref{selfa})  yields
\begin{eqnarray}\label{azx}
\frac{\delta}{U} &\simeq& \frac{\Delta}{2|\Gamma\cos(\phi/2)-\delta|} \\
&\times& \nonumber
\left( \frac{2\Gamma\cos(\phi/2) \cos^{-1}\big|\frac{
\Gamma\cos(\phi/2)-\delta}{\Delta}\big|}{\pi \sqrt{\Delta^2-\left[
\Gamma\cos(\phi/2)-\delta\right]^2}} -\frac{ \delta}{\Delta}\right).
\end{eqnarray}
The interaction correction
to the Josephson current now follows from Eq.~(\ref{jos}),
\[
\delta I\simeq \frac{2e\delta}{\hbar}\tan(\phi/2)\left(
\frac{1}{2} f_1(\phi)\,{\rm sgn}\cos{(\phi/2)} -\frac{\delta}{U}\right).
\]
with
\[
f_1(\phi)=\frac{1+2 (\Gamma/\Delta) |\cos(\phi/2)|}
{[1+(\Gamma/\Delta)|\cos(\phi/2)|]^2}=1+O(\Gamma^2/\Delta^2).
\]
In the extreme limit $\Gamma/\Delta\to 0$,
Eq.~(\ref{azx}) yields $\delta=
(U/2){\rm sgn}\cos(\phi/2)$, and then $\delta I=0$.
Inspection of Eq.~(\ref{azx}) for finite $\Gamma/\Delta\ll 1$
shows however that $|\delta|<U/2$ for $\phi \ne \pi$.
As a result, the interaction current to the Josephson current
for $\Gamma\ll\Delta$ turns out to be positive, albeit
numerically small. Using  Eq.~(\ref{azx}) we find
\[
\delta I\simeq \frac{eU}{\hbar \pi}\frac{\Gamma}
{\Delta}\sin(\phi/2)\,{\rm sgn}\cos{(\phi/2)}.
\]
We believe that this effect is related to
the enhancement of the current at finite bias $V$ in the regime
$\Gamma\ll \Delta$ discussed in Sec.~\ref{sec3}.

On the other hand, for $\Gamma\gg \Delta$, Eq.~(\ref{selfa}) is solved by
$\delta\simeq U\Delta \ln(\Gamma/\Delta)\cos(\phi/2)/(\pi\Gamma)$,
and the lowest-order interaction correction to
the Josephson current is
\begin{eqnarray*}
\delta I &\simeq& \frac{2e\delta}{\hbar}\tan(\phi/2)
\left( \frac{\Delta}{2\Gamma} f_2(\phi)\,
{\rm sgn}\cos(\phi/2)
-\frac{\delta}{U}\right)
\end{eqnarray*}
with $f_2(\phi)=1+\cos^2(\phi/2) /4 $.
Due to the large $\ln(\Gamma/\Delta)$ factor appearing now in $\delta$,
the Josephson current will in general be decreased by interactions for
$\Gamma\gg \Delta$.

We therefore find the same qualitative picture as for the nonequilibrium
current in Sec.~\ref{sec3}: The equilibrium Josephson current can also
be slightly increased by weak repulsive interactions for weak hybridization,
$\Gamma/\Delta\ll 1$, but is decreased in the opposite limit.
However, the increase of the Josephson current
for weak hybridization turns out to be much smaller than
for the corresponding nonequilibrium current.

\end{appendix}


\begin{thebibliography}{99}

\bibitem{vandam} J. van Dam, Yu.V. Nazarov, E.P.A.M. Bakkers, S. De Franceschi,
and L.P. Kouwenhoven, Nature {\bf 442}, 667 (2006).

\bibitem{tube} A.Yu. Kasumov {\sl et al.}, Science {\bf 284}, 1508 (1999);
A. Morpurgo, J. Kong, C.M. Marcus, and H. Dai, Science {\bf 286}, 263 (1999);
M.R. Buitelaar, T. Nussbaumer, and C. Sch\"onenberger, Phys. Rev. Lett.
{\bf 89}, 256801 (2002);
J.-P. Cleuziou, W. Wernsdorfer, V. Bouchiat, T. Ondarcuhu, and M. Monthioux,
Nature Nanotechnology {\bf 1}, 53 (2006).

\bibitem{buitelaar} M.R. Buitelaar, W. Belzig, T. Nussbaumer,
B. Babic, C. Bruder, and C. Sch\"onenberger, Phys. Rev. Lett. {\bf 91}, 057005 (2003).

\bibitem{fulleren} A.Yu. Kasumov, K. Tsukagoshi, M. Kawamura, T. Kobayashi,
 Y. Aoyagi, K. Senba, T. Kodama, H. Nishikawa, L. Ikemoto,
K. Kikuchi, V.T. Volkov,
Yu.A. Kasumov, R. Deblock, S. Gu{\'e}ron, and H. Bouchiat,
Phys. Rev. B {\bf 72}, 033414 (2005).

\bibitem{jorgensen} H.I. Jorgensen, K. Grove-Rasmussen, T. Novotny,
K. Flensberg, and P.E. Lindelof, Phys. Rev. Lett. {\bf 96}, 207003 (2006).

\bibitem{lieber} J. Xiang, A. Vidan, M. Tinkham,
 R.M. Westervelt, and C.M. Lieber,
Nature Nanotechnology {\bf 1}, 208 (2006).

\bibitem{jarillo} P. Jarillo-Herrero, J.A. van Dam, and L.P. Kouwenhoven,
Nature {\bf 439}, 953 (2006).

\bibitem{christian} A. Eichler, M. Weiss, S. Oberholzer,
C. Sch\"onenberger, A. Levy Yeyati, J.C. Cuevas,  and A. Martin-Rodero,
Phys. Rev. Lett. {\bf 99}, 126602 (2007).

\bibitem{lindelof} T. Sand-Jespersen, J. Paaske, B.M. Andersen,
K. Grove-Rasmussen, H.I. Jorgensen, M. Aagesen, C.B. Sorensen,
P.E. Lindelof, K. Flensberg, and J. Nygard,
Phys. Rev. Lett. {\bf 99}, 126603 (2007).

\bibitem{mar} E.N. Bratus, V.S. Shumeiko, and G. Wendin, Phys. Rev. Lett. {\bf 74}, 2110 (1995);
D.V. Averin and D. Bardas, {\sl ibid.} {\bf 75}, 1831 (1995);
J.C. Cuevas, A. Martin-Rodero, and A. Levy Yeyati,
Phys. Rev. B {\bf 54}, 7366 (1996).


\bibitem{ingerman}
\AA. Ingerman, G. Johansson,
V.S. Shumeiko, and G. Wendin, Phys. Rev. B {\bf 64}, 144504 (2001);
J. Lantz, V.S. Shumeiko, E.N. Bratus, and G. Wendin,
{\sl ibid.} {\bf 65}, 134523 (2002).

\bibitem{cuevas} A. Levy Yeyati, J.C. Cuevas, A. Lopez-Davalos,
and A. Martin-Rodero, Phys. Rev. B {\bf 55}, R6137 (1997).

\bibitem{wendin} G. Johansson, E.N. Bratus,
V.S. Shumeiko, and G. Wendin, Phys. Rev. B {\bf 60}, 1382 (1999).

\bibitem{siano}
L. Glazman and K.A. Matveev,
JETP Lett. {\bf 49}, 659 (1989); A.V. Rozhkov and D.P. Arovas,
Phys. Rev. Lett. {\bf 82}, 2788 (1999); E. Vecino, A. Martin-Rodero, and
A. Levy Yeyati, Phys. Rev. B {\bf 68}, 035105 (2003);
F. Siano and R. Egger, Phys. Rev. Lett. {\bf 93}, 047002 (2004);
M.S. Choi, M. Lee, K. Kang, and
W. Belzig, Phys. Rev. B {\bf 70}, 020502(R) (2004);
C. Karrasch, A. Oguri, and V. Meden, {\sl ibid.} {\bf 77}, 024517 (2008).

\bibitem{kang} K. Kang, Phys. Rev. B {\bf 57}, 11891 (1998);
Physica E {\bf 5}, 36 (1999); S.Y. Liu and X.L. Lei, Phys. Rev. B {\bf 70},
205339 (2004).

\bibitem{avishai} Y. Avishai, A. Golub, and A.D. Zaikin,
Phys. Rev. B {\bf 63}, 134515 (2001).

\bibitem{slave}  F.S. Bergeret, A. Levy Yeyati, and A. Martin-Rodero,
Phys. Rev. B {\bf 74}, 132505 (2006); Y. Avishai, A. Golub, and
A.D. Zaikin, {\sl ibid.} {\bf 67}, 041301(R) (2003).

\bibitem{alfredo} A. Levy Yeyati, A. Martin-Rodero, and E. Vecino,
Phys. Rev. Lett. {\bf 91}, 266802 (2003).

\bibitem{normal} T. Fujii and K. Ueda, J. Phys. Soc. Jpn.
{\bf 74}, 127 (2005).

\bibitem{hamasaki} M. Hamasaki, Condensed Matter Physics {\bf 10}, 235 (2007).

\bibitem{eg}
R. Egger and A.O. Gogolin, cond-mat/0712.0750 (to appear in Phys. Rev. B).

\bibitem{zazu} A. Zazunov, R. Egger, C. Mora, and T. Martin,
Phys. Rev. B {\bf 73}, 214501 (2006).

\bibitem{zarand}
B. Horv{\'a}th, B. Lazarovits, O. Sauret, and  G. Zar{\'a}nd,
cond-mat/0712.0296.

\bibitem{baym} G. Baym and L.P. Kadanoff, Phys. Rev. {\bf 124}, 287 (1961);
{\sl ibid.} {\bf 127}, 1391 (1962).

\bibitem{hershfield} S. Hershfield, J.H. Davies, J. W. Wilkins, Phys. Rev. B
{\bf 46}, 7046 (1992).

\bibitem{alf3}
A. Levy Yeyati, J.C. Cuevas, and A. Martin-Rodero, Phys. Rev. Lett. {\bf 95}, 056804 (2005).


\end{thebibliography}
\end{document}